\documentclass[%
 reprint,
 superscriptaddress,
showpacs,
 amsmath,amssymb,
 aps,%onecolumn,
 twocols,
]{revtex4-1}
\usepackage{graphicx,color}% Include figure files
\usepackage{xcolor}
\usepackage{dcolumn}% Align table columns on decimal point
\usepackage{bm}% bold math
\usepackage{booktabs} %professional-like tables
\usepackage[separate-uncertainty=true]{siunitx}
\usepackage{multirow}

\begin{document}
\title{Chirality modifies the interaction between knots}

\author{Saeed Najafi}
\affiliation{Max Planck Institute for Polymer Research, Ackermannweg 10, 55128 Mainz, Germany}
\author{Luca Tubiana}
\affiliation{University of Vienna, Computational Physics Group, Sensengasse 8, 1090, Vienna, Austria}
\author{Rudolf Podgornik}
\affiliation{University of Ljubljana, Faculty for Mathematics and Physics, Jadranska 19, 1000, Ljubljana}
\author{Raffaello Potestio}
\email{potestio@mpip-mainz.mpg.de}
\affiliation{Max Planck Institute for Polymer Research, Ackermannweg 10, 55128 Mainz, Germany}

\date{\today}

\begin{abstract}
In this study we consider an idealization of a typical optical
	tweezers experiment involving a semiflexible double-knotted polymer, with
	steric hindrance and persistence length matching those of dsDNA in high salt
	concentration, under strong stretching. Using exhaustive Molecular Dynamics simulations we show that
	not only does a double-knotted dsDNA filament under tension possess a free
energy minimum  when the two knots are intertwined, but also that the depth of
this minimum depends on the relative chirality of the two knots. We rationalize
this dependence of the effective interaction on the chirality in terms of a
competition between chain entropy and bending energy.

\end{abstract}

\maketitle

The study of physical knots in polymers is an important emerging topic
in biophysics and soft matter in general. Since the original conjecture that knots should be ubiquitous
in sufficiently long chains~\cite{Delbruck_knot_62,Frisch:JACS:1961}, later proved by
Sumners and Whittington~\cite{SumnersWhittington1988}, knots have been
found or tied in a variety of biopolymers, from DNA~\cite{ArsuagaPNAS2002,ArsuagaPNAS2005,MarenduzzoPNAS2009}
 to proteins~\cite{Taylor:Nature:2000,Virnau:PLoScb:2006,King:JMB:2007,potestio:2010:PLoS,Skrbic:PLoScb:2012,Beccara:PLoScb:2013,najafi_jcp2015}
and even actin filaments~\cite{Arai:1999:Nature}, and have been shown to have a large impact
on the biological function of proteins
and DNA~\cite{Bates:DNA,meluzzi2010biophysics}. More recently, several
studies have shed light on the relevance of knots also in nanotechnological
applications~\cite{Arai:1999:Nature, Micheletti_PhRep2011,Micheletti:SoftMatt:2012,Rosa:PRL:2012,Coluzza:PRL:2013}.

Physical knots appear and diffuse spontaneously along polymer chains~\cite{BenNaim:prl:2001,Virnau:JACS:2005,Tubiana:Macromol:2013A,Matthews:EPL:2010,Volodogskii:BioJ:2006,Huang:JPCA:2007}, on which they acquire a metastable tightness~\cite{Grosberg:PRL:2007,Dai:Macromol:2014,Dai:PRL:2015},
and can affect structural and dynamical properties like
their radius of gyration~\cite{desCloizeux:1981:J-Phys-Lett,Moore:2004:PNAS,Orlandini:2010:PRE,Mansfield:JCP:2010},
tensile strength~\cite{Saitta_et_al:1999:Nature,Arai:1999:Nature},
diffusion constants~\cite{Stasiak:1996:Nature:8906784, Weber_et_al_2006_Biophys_J},
and translocation dynamics through a pore~\cite{Rosa:PRL:2012}.
%Conversely,
%knots probabilities, sizes and diffusion coefficient depends on the physics of
%the polymers in which they are tied. These dependencies can be used to gain insight 
%into the physics of polymer systems, as exemplified by the use of knot spectra 
%to study the arrangement of DNA inside bacteriophages capsids~\cite{}, or possibly
%to devise new strategies of detecting different knots in single-molecule stretching
%experiments, based on their different diffusion coefficient and the sensible dendence on 
%the low force regime on knot sizes. 

Most studies up to now focused on the properties of single knots, investigating how knot occurrence probability~\cite{Koniaris:prl:1991,Orlandini:RevModPhys:2007},
	size~\cite{Millett:Macromol:2005,Marcone:PRE:2007,Mansfield:JCP:2010}, and dynamical properties depend on physical characteristics of the system under study, such as polymer thickness~\cite{Rybenkov:1993:PNAS,Shimamura:JPA:2000}, confinement~\cite{ArsuagaPNAS2005,MarenduzzoPNAS2009,Micheletti:SoftMatt:2012,Micheletti:MacroLett:2014,Poier:Macromol:2014}, stretching force~\cite{Farago:2002:EPL,Caraglio:PRL:2015,Huang:JPCA:2007,Matthews:EPL:2010},
crowding and solution density~\cite{Kim:Macromol:2004,Rosa:Macromol:2011,dAdamo:Macromol:2015}. Polymers, though, can
host multiple knots. These, referred to as \emph{composite knots}, are actually the most probable type of knots in the case of long
polymers~\cite{SumnersWhittington1988,Diao:JKTR:1994}.  The study of
composite knots is of great interest since the presence of interactions among
their prime components may alter the overall properties of the polymer.
For example, knot colocalization on a stretched polymer may significantly diminish its
tensile strength with respect to a chain with a series of
localized, non-intertwined prime knots.

\begin{figure}
\includegraphics[width=\columnwidth]{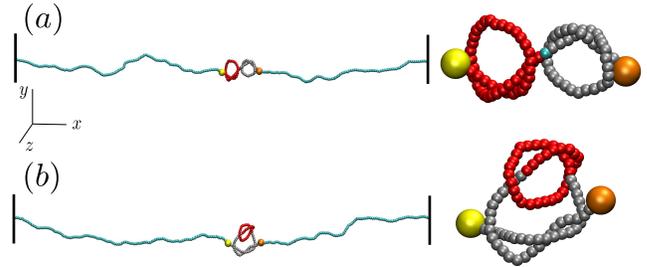}
\caption{\label{fig:schematic_state}  Snapshots from simulations of a chain
	with bending rigidity $\kappa = 20 k_BT$ containing two $3_1$ knots of
	different chirality.  The shortest knotted portions of isolated prime
	components (zoomed in the right part of the figure) are highlighted in red
and grey.  Yellow and orange beads, whose sizes were artificially increased for clarity, mark the beginning and end of the composite knots, respectively.}
\end{figure}

Since prime knots tied on polymers in solutions or under weak mechanical stretching are weakly localized, their length growing sublinearly with
the length of the polymer~\cite{Farago:2002:EPL,Millett:Macromol:2005,Marcone:PRE:2007,Mansfield:JCP:2010,Caraglio:PRL:2015},
when tied on the same polymer they are expected, in the thermodynamic limit of infinitely long
chains, to be statistically independent.  In fact, computational
studies have shown that in the thermodynamic limit the characteristic exponent,
relating the configurational entropy of a polymer ring to its contour length,
can be written as in the case of an unknotted ring augmented by the number of knots that are
present, and that the knotting probability of composite knots factorizes into
that of their prime knot components~\cite{Baiesi:JSM:2010,JPSJ_Tsurusaki_1995}.
However, in finite-size polymers the situation can be quite different. In this
case the size of knots is non-negligible and therefore they can not be mapped
onto independent point-like decorations~\cite{Zandi:ArXiv:2003}.  On
the contrary, knots can be expected to be intertwined, so that one is inside
another, and to interact with each other.  Dommersnes ~\emph{et al.} have shown
that knots tied on a short unscreened charged ring become as tight as possible
and maximize their relative distances along the
ring~\cite{Dommersnes:PRE:2002}, while Virnau and coworkers, simulating a
double-knotted stretched dsDNA chain, have shown that knots can become
intertwined in a minimum free energy configuration~\cite{Trefz:PNAS:2014}. Furthermore, a recent study by one of
us~\cite{Tubiana:PRE:2014} reported that knot size remains an important factor
in freely jointed rings of up to thousands of bonds, where the probability of
finding intertwined knots remains significant. These results suggest that
finite size effects are relevant in most biological phenomena and
nanotechnological applications involving stiff polymers such as dsDNA, and that
indeed the overall effect of knots may depend on some complex interaction
between them.

To shed further light on the interactions between knots along a finite-size
polymer, we consider an idealization of a typical optical tweezers
experiment~\cite{Trefz:PNAS:2014, Farago:2002:EPL,Caraglio:PRL:2015} in which a
semiflexible double-knotted chain is stretched between two impenetrable
walls, describing the effect of  tweezer confinement on the terminal
polystyrene beads (see Fig. 1).  The separation of the chain termini 
	is such that the knotted polymer is in the high-force stretching regime in which 
knots are strongly localized, their size showing only small fluctuations around an 
average size which scales solely with the applied force~\cite{Farago:2002:EPL,Caraglio:PRL:2015}.
	Using exhaustive Molecular Dynamics (MD) simulations of polymers containing different pairs of prime knots, we show that
not only does a double-knotted dsDNA filament under tension possess a free energy
minimum  when the two knots are intertwined~\cite{Trefz:PNAS:2014}, but also that
the depth of this minimum depends on the relative chirality of the two knots.
We show that this dependence of the effective interaction on the chirality  originates from an interplay between chain entropy and bending energy, which is dominated by the latter.

The dsDNA is modeled as a chain of $N=300$ beads of diameter $\sigma$ connected by FENE
(finitely extensible nonlinear elastic)
bonds~\cite{kremer_grest_1986,Kremer:1990:JCP}. A Weeks-Chandler-Anderson (WCA)
potential enforces the excluded volume, and a harmonic bending energy term
induces a finite persistence length. The chain is stretched along the $X$ axis,
and its termini are kept fixed in correspondence of two repulsive walls,
modeled as WCA potentials, which prevent the knots from untying.
The total potential energy of the system is thus:
\begin{eqnarray}
	U_{tot} = U_{\rm \small WCA} + U_{\rm \small FENE} +  U_{\rm \small bending} + U_{\rm \small walls}
\label{eqn:ham}
\end{eqnarray}
where the WCA potential is taken as:
\begin{equation}\label{eqn:potentials}
\begin{split}
&U_{\rm \small WCA} = \frac{1}{2} \sum_{(i,j),j\not=i}^N V(d_{i,j})\\
&V(r) = \left\{
  \begin{array}{l}
  4 \epsilon \left[ \left(\frac{\sigma}{r}\right)^{12} - \left(\frac{\sigma}{r}\right)^{6} + \frac{1}{4} \right]\ \mbox{for}\ r \le 2^{1/6} \sigma\\
  0\ \ \mbox{otherwise}\\
  \end{array}\right..
\end{split}
\end{equation}
The WCA strength $\epsilon =1k_BT$ and the characteristic length scale $\sigma$ are
taken as the energy and length units, respectively. All other dimensional
quantities are expressed in terms of reduced units defined through $\epsilon$,
$\sigma$ and the bead unit mass $m$. Time is measured in the MD
time units $\tau_{MD} = \sigma\sqrt{m/\epsilon} = 1$. The FENE potential reads:
\begin{eqnarray}
  U_{\rm \small FENE} = - \sum_{i=0}^{N-2} \frac{{\kappa_{fene}}}{2} \left( \frac{R_0}{\sigma} \right)^2 \ln \left[ 1 - \left( \frac{\vert\vec{u}_i\vert}{R_0} \right)^2 \right]
\end{eqnarray}
where $\vec{u}_i \equiv \vec{r}_{i+1}-\vec{r}_{i}$ is the vector pointing to
the bead $i+1$ from bead $i$, and $\vert\vec{u}_i\vert$ is thus the distance of
the bead centers $i$ and $i+1$. The values of the maximum bond length $R_0 =
1.5\sigma$ and the FENE interaction strength $\kappa_{fene} = 30 \epsilon$ are
the customary ones for the Kremer-Grest model \cite{kremer_grest_1986}. The
harmonic bending potential is taken in its standard form:
\begin{equation}
U_{\rm \small bending} = \sum_{i=1}^{N-1} {{\kappa} \left(1 - \frac{\vec{u}_i\cdot\vec{u}_{i+1}}{|\vec{u}_{i}| |\vec{u}_{i+1}|} \right)}
\end{equation}
where $\kappa = 20 k_B T$ is the bending stiffness of the chain, inducing
a persistence length $l_p=20\sigma$. Setting $\sigma=2.5$ nm gives us
$l_p=50$ nm, typical of dsDNA in high monovalent salt concentrations~\cite{Rybenkov:1993:PNAS}.

 The potential of Eq. \ref{eqn:ham} is used to perform underdamped MD
 simulations in an implicit solvent with a Langevin thermostat and time step
 $\Delta t = 0.01 \tau_{MD}$, with the friction self-correlation time
 $\tau_{frict} = 10^3\tau_{MD}$.

To assign a topological state to subsections of the chain, we used the {\it
Minimally Interfering Closure}~\cite{Tubiana:PTPS:2011}. We define a knotted
portion of the chain as the shortest segment featuring a specific knotted
topology upon closure according to the Alexander polynomial \cite{Note0}. With some abuse of language, we
will refer to such portions as ``knots" in what follows.  By applying this
procedure to composite knots, we are able to identify both the chain portion
hosting the whole composite knot, as well as those hosting its ``isolated''
prime components.  Following ref.~\cite{Tubiana:PRE:2014}, we consider a prime
component to be isolated when it can be excised, and its ends joined,  without
at the same time untying the second knot, see Fig~\ref{fig:schematic_state}.

Six topologically different setups have been investigated, namely: $(3_1^+
3_1^+)$, $(3_1^+ 3_1^-)$, $(4_1 3_1^+)$, $(4_1 3_1^-)$, $(5_1^- 3_1^-)$,
$(5_1^- 3_1^+)$. The chirality of each prime knot has been established
	using the writhe, that is, the sum of the signed crossings of the knot in
	its reduced diagram. $+$ and $-$ superscripts refer to \emph{right} and \emph{left}
	handedness according to the right-hand rule convention~\cite{Micheletti_PhRep2011,Kauffman:TransAMS:1990}.
	In our setup, ($++$) and ($--$) composite knots are related by a mirror transformation; therefore,
	we do not attach any importance to the overall chirality of the composite knot,
	but only on the relative chirality of its prime components.  Since no cross-passage is allowed in our simulations, the chirality of all knots is preserved during the
simulations. In all setups, the termini of the chain were kept fixed at a distance
of $L=205\sigma$, corresponding to forces of about $1$-$4$ pN at $T=300 K$, depending on the
knot complexity, applied on both termini. At these forces, corresponding to a strong stretching regime~\cite{Caraglio:PRL:2015},
the knot lengths show relatively small fluctuations around their average
values, as reported in Table~\ref{tab:LK}. For each of the six topologies under examination, $40$
independent simulations were performed, each consisting of an initial
equilibration phase of $2\times{10^7}$ $\tau_{MD}$ and a production phase of
$2\times{10^9}$ $\tau_{MD}$.

\begin{table*}[htp]
\centering
\begin{tabular}{|l|l|l|l|l|}
\hline
\multirow{2}{*}{} & \multicolumn{2}{l|}{$~~~~~~~$Separated} & \multicolumn{2}{l|}{$~~~~~~$Intertwined} \\ \cline{2-5}
                                    &   $~~~~$$l_k^{3_1}$       &   $~~~~$$l_k^{other}$                 &$~~~~$$l_k^{incl}$  & $~~~~$$l_k^{comp}$      \\ \hline
$5_1^-$ \# $3_1^-$                  &    $29.5\pm3.1$       & $50.5\pm4.3$ &    $30.3\pm4.7$   &    $72.4\pm3.7$       \\ \hline
$5_1^-$ \# $3_1^+$                  &    $29.6\pm3.1$       & $50.6\pm4.3$ &    $33.9\pm5.2$   &    $72.9\pm3.7$       \\ \hline
$4_1$ \# $3_1^+$                    &    $32.9\pm3.6$       & $44.5\pm4.0$ &    $30.4\pm3.3$   &    $72.6\pm3.8$       \\ \hline
$4_1$ \# $3_1^-$                    &    $32.9\pm3.7$       & $44.5\pm4.1$ &    $30.4\pm3.3$   &    $72.6\pm3.8$       \\ \hline
$3_1^+$ \# $3_1^+$                  &    $35.5\pm2.2$       &                          &    $33.0\pm3.7$   &    $67.2\pm4.1$       \\ \hline
$3_1^+$ \# $3_1^-$                  &    $35.7\pm2.3$       &                          &    $34.7\pm3.9$   &    $67.0\pm4.0$       \\ \hline
\end{tabular}
\caption{Average knot lengths for different topologies. In the ``Separated" columns
are reported the knot lengths for the prime components when they are not intertwined.
The ``Intertwined" columns report the average lengths of the isolated prime component
that has been included (labeled $l_k^{incl}$) and of the whole composite knot in an
intertwined configuration (labeled $l_k^{comp}$). 
In those cases where the $4_1$ is entwined by the $3_1$, constituting the $8\%$ of the
intertwined configurations for this topology, the size of the $4_1$ is $42.5\pm3.7$,
and the length of the whole composite knot is $70.1\pm3.6$. We remark that the knots 
	under investigation here are quite tight, with lengths almost half (55\%) those of the knots
studied in ref.~\cite{Trefz:PNAS:2014} in the case of the $3_1\#4_1$ systems. }
\label{tab:LK}
\end{table*}

\begin{figure}[ht]
\begin{center}
\includegraphics[width=0.66\columnwidth]{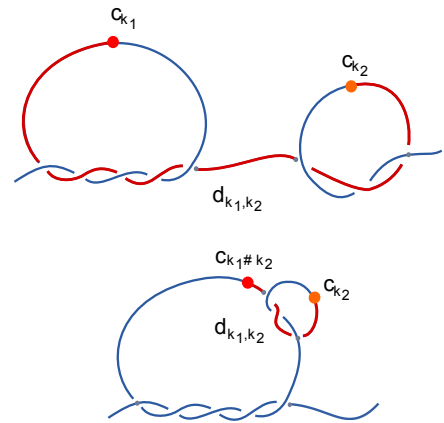}
\caption{\label{fig:knot:distance} Schematics of the collective order parameter
	$D$ measuring the linear distance between two prime knots, in this case a
	$5_1$ and a $3_1$. When both prime components are isolated, the order
	parameter is given by $D=c_{k_2}-c_{k_1}$, where
	$c_{k_i}=(e_{k_i}+s_{k_i})/2$ is the center of knot $i$ on the chain. Here
	$e_{k_i}$ and $s_{k_i}$ stand for the last and the first bead of the $i$-th
	isolated knotted portion. $k_1$ is always taken to be the most complex knot,
	in this case the $5_1$. When the knots are intertwined, we identify the
	center of the swollen knot with the center of the whole composite
	knot, $c_{k_1\#k_2}$.
}
\end{center}
\end{figure}

As a first case we investigate the $3_1^\pm5_1^-$ system, taking into account
two different chiralities of the trefoil knot: $3_1^+$ and $3_1^-$. The
fractions of intertwined states in the $3_1^+$ and $3_1^-$ cases are 0.585 and
0.447, respectively. In both setups the largest knot, the $5_1$, swells up to
let the $3_1$ knot in. The frequency with which the trefoil enters or exits the $5_1$ is
$7.6\times10^{-9}\ \tau_{MD}^{-1}$ for the $5_1^-3_1^-$ ($--$) pair, and $13.2\times10^{-9}
\ \tau_{MD}^{-1}$ for the $5_1^-3_1^+$ ($-+$) pair \cite{Note1}.
The data in Table~\ref{tab:LK} show that the length of separate prime components is independent of their
relative chirality for all topologies under study. This allows us to introduce a collective descriptor, or an {\it order
parameter}, $D$, defined as the oriented distance between the knot centers.
This is measured as the number of chain beads from the center of the $5_1$ knot to the
center of the $3_1$ knot, $D=c_{3_1} -c_{5_1}$.  A similar definition can be provided for the
configurations in which the two knots are intertwined and the knot
identification algorithm allows us to identify only one prime component,
the one which has been entwined by the other knot. In those cases we identify
the center of the swollen  knot with the center of the whole composite knot in the expression for $D$ \cite{Trefz:PNAS:2014}.
 Therefore, $D=0$  in those configurations in which the two knots are intertwined and the innermost knot is located exactly at the center of the outermost knot.
 A schematic representation of this collective descriptor is provided in Fig.~\ref{fig:knot:distance}.

\begin{figure}[]
\begin{center}
\includegraphics[width=\columnwidth]{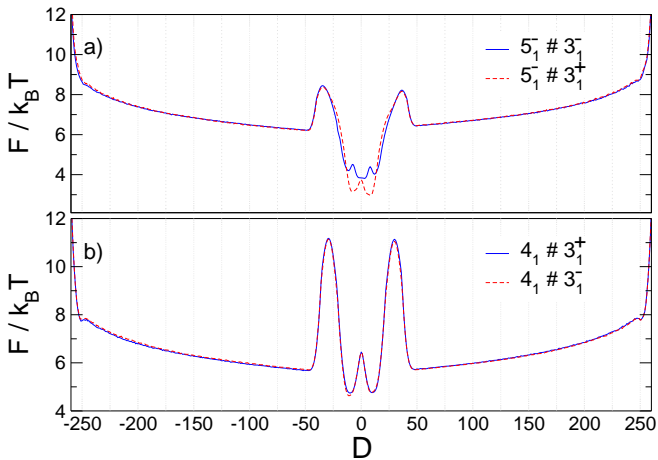}
\caption{\label{fig:5_1} (a) Free energy, $F(D)$, as a function of the linear distance between the
knots for $5_1^- 3_1^-$ (blue solid line) and $5_1^- 3_1^+$ (red dashed line). (b) Same as
in (a) for the knots $4_1 3_1^+$ (blue solid line) and $4_1 3_1^-$ (red dashed line). Note
that in this latter case the two quantities coincide.  The free energies for the $4_1 3_1^\pm$
reported in panel (b) can be decomposed to distinguish the cases in which
the $4_1$ intakes the $3_1$ in and those in which the opposite happens; no substantial
difference depending on the relative chirality is to be observed (data not shown).}
\end{center}
\end{figure}

By counting the relative number of MD configurations for which the knot
components are separated by a given distance $D$ we can obtain the probability distribution $P(D)$ and, correspondingly, the free energy $F(D) = - k_B T \ln P(D)$. The latter is reported for the two $3_1^\pm 5_1^-$ systems in Fig.~\ref{fig:5_1}a.

From the profiles in Fig.~\ref{fig:5_1}a we observe that $F(D)$ increases
with increasing $|D|$, a behaviour that can be intuitively attributed to the
entropic cost of placing two knots at large distance on a long, yet finite
chain~\cite{Zandi:ArXiv:2003}.  Consistent also with previous
observations~\cite{Trefz:PNAS:2014}, for small values of $|D|$ we detect two
barriers and two minima in the free energy,  corresponding to configurations in
which the two knots are intertwined.  Most interestingly and unexpectedly, we
observe that the depth of these minima depends on the {\it relative chirality}
of the knots. When the two components have opposite chirality, the
corresponding free energy minimum is $\sim 1 k_BT$ deeper than for the system
in which the two chiralities are identical.

This conclusion is reinforced by simulations performed on a $3_1^\pm 4_1$
composite knot, where one of the knots, the $4_1$, is achiral. The free energy
in this case, reported in Fig. \ref{fig:5_1}b, does not depend on the chirality of the $3_1$ component, as it is indeed
expected since there are no other chiral entities in the setup. In the simulations the intertwined
states with the $4_1$ including the trefoil are the most probable, making for the 92\% of
observed configurations~\cite{Note2}.

\begin{figure}[t!]
\begin{center}
\includegraphics[width=\columnwidth]{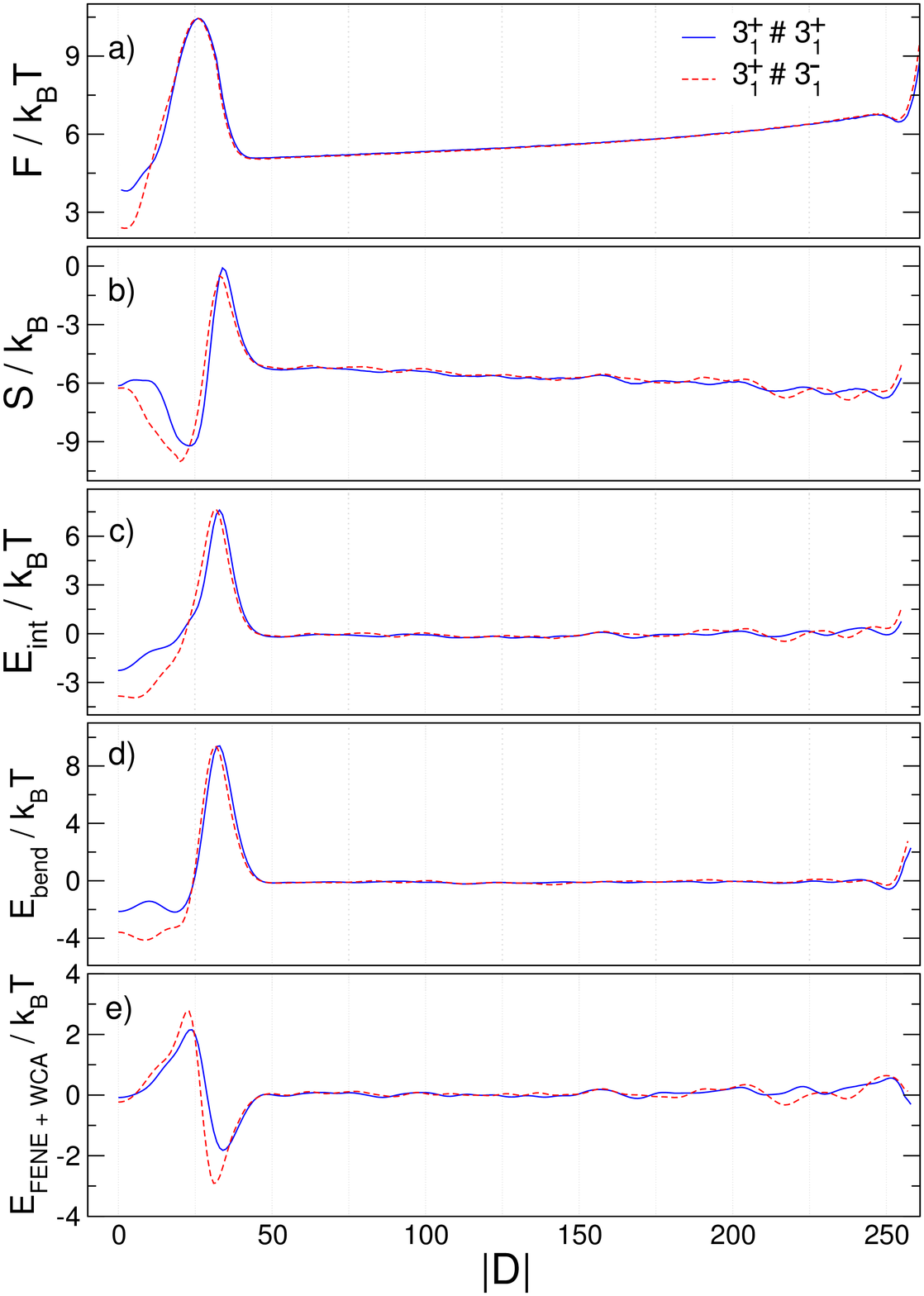}
\caption{Free energy of the $3_1 3_1$ system with same and opposite
relative chirality, as a function of the separation $D$ between the two knots,
decomposed in their different energetic and entropic contributions. We report the free energy $F(D)$ (a), the entropy $S$ (b),
and the internal potential energy $E_{int}$ (c). The entropy is obtained through
the relation $F = E_{int} - T S$. In panels (d) and (e) we show how the internal
potential energy is distributed, respectively, between the bending term and the
remaining two contributions, namely the FENE and WCA potentials.
See the main text for a discussion of the results.\label{fig:3_1}}
\end{center}
\end{figure}

To understand if the sole relative chirality of two knots  can mark a
difference in their preference to stay intertwined or to separate along a chain
under tension, we consider a system composed by two otherwise identical trefoil
knots. In this case we use as order parameter the absolute value of $D$, since
when the two trefoil knots have the same chirality they become effectively
indistinguishable.  The free energy profiles, Fig.~\ref{fig:3_1}a, corroborate
that also in this case the system with two knots having opposite chiralities
has a lower free energy minimum when the two components are intertwined.
We note here that the presence of the two repulsive walls may impact the
	free energy profiles, since our parameter $D$ does not distinguish the
	whether the knots are near the walls or far from them.  In order to rule out
	possible distortions due to the interactions between knots and walls, the
	free energy profiles have been also computed excluding all those
	configurations in which the knots were separated from the wall by a distance
	lower than $2 l_p$. The resulting profiles (data not shown) are perfectly
consistent with the ones reported in Figs. \ref{fig:5_1} and \ref{fig:3_1},
thus validating the robustness of the observed behavior.

Further insight into the cause of this effect can be obtained by
analyzing the size of the intertwined configurations. As reported in
Table~\ref{tab:LK}, we find that for all investigated topologies the length of
the composite knot in the intertwined state, $l_{comp}$, does not depend on the
relative chirality of the prime components. Interestingly, we observe that both for the
$3_1^\pm 5_1^-$ and for the $3_1^\pm 3_1^-$ systems the length of the
nested, isolated prime components in the intertwined state is slightly larger in
the ($+-$) case \cite{Note3}.

We proceed to separate the free energy $F(D)$ in its energetic and
entropic components,  by first computing the average internal (potential)
energy of the configurations, and subsequently obtaining the entropy through
the standard relation $F=E_{int} - TS$. The results, reported in Fig.~\ref{fig:3_1}a-c, show two interesting features. First, they confirm that the observed increase of $F$ with $|D|$ when the two knots are separated is purely entropic. Secondly, and more importantly, they show that the differences we observe in the free energies of the ($+-$) and ($++$) systems originate from a complex interplay of internal energy and entropy. Specifically, the entropic contribution is higher for the ($++$) case but is not high enough to overcome the energetic contribution favoring the ($+-$) knot.
The potential energy can be further decomposed into its main
components: steric hindrance, bond extension, and bending energy.  The data
presented in Fig.~\ref{fig:3_1}d-e clearly show that while all
other energetic contributions are similar, the bending energies of the ($+-$)
and ($++$) systems differ significantly at the position of the minimum of
$F(D)$ by the same amount, $\sim 1.5 k_BT$. The same qualitative result holds
also for the $3_1^\pm5_1^-$ topologies (data not shown).

Given the observed competition between bending energy and entropy in the intertwined state, with the ($+-$) system showing a lower bending energy  but losing more entropy than the ($++$) system, it is tempting to ascribe the difference in their free energy profile to a significantly different arrangement of the nested knot within the hosting knot in the two setups. However, further analyses and simulations are required to elucidate  the exact mechanics underlying the chiral contribution, which is the object of an ongoing study.

Summing up, we have shown that a double knotted semiflexible polymer chain
under strong stretching possesses a free energy minimum when the two knots are
intertwined, showing that relatively tight knots can still pass through each other,
and also that the depth of this minimum depends on the relative
chirality of the two knots. In order to observe this effect, both knots must be
chiral, with the knots of opposite chirality displaying a higher preference to
stay intertwined.  Furthermore, we showed that the major player in the chirality
effect is the bending energy of the chain, which, we recall, is set to the
characteristic persistence length of dsDNA.

The question then emerges, as to what are the specific features of the bending
energy that would engender a chirality effect in knot interaction, and whether
the stretching of the chain enters in some way as a significant constraint. If
the chirality effect, described above, turns out to be robust and observable in
a broad range of parameters, one can speculate further as to its importance in
particular for chiral polymers such as dsDNA.

\acknowledgments
	L. Tubiana and R. Podgornik acknowledge support from the Slovenian Agency
	for Research and	Development (ARRS grant No J1-4134). L. Tubiana acknowledges
	also support from the Mahlke-Oberman Stiftung and the European Union's Seventh
	Framework Programme for research, technological development and demonstration
	(grant No 609431). S. Najafi and R. Potestio are thankful to M. Heidari and R. Menichetti
	for an attentive reading of the manuscript and useful comments.

%\section{Introduction}\label{sec:intro}
%%%%%%%%%%%%%%%%%%%%%%%%%%%%%%%%%%%%


\end{document}